# Label free optical transmission tomography for biosystems: intracellular structures and dynamics

OLIVIER THOUVENIN,[1] SAMER ALHADDAD,[1] VIACHESLAV MAZLIN,[1] MARTINE BOCCARA[2,3] AND CLAUDE BOCCARA[1*]

[1]*Institut Langevin, ESPCI Paris, PSL University, CNRS, 1 rue Jussieu, 75005 Paris, France*
[2]*Ecole Normale Supérieure, PSL University, Institut de Biologie de l'Ecole Normale Supérieure (IBENS), CNRS UMR 8197, INSERM U1024, 46 rue d'Ulm, F-75005 Paris, France*
[3]*Institut de Systématique, Evolution, Biodiversité (ISYEB), Muséum National d'Histoire naturelle, CNRS, SU, EPHE, UA, 57 rue Cuvier CP50, 75231 Paris Cedex 05, France*
*\*claude.boccara@espci.fr*

**Abstract:** There is an increasing need for label free methods that could reveal intracellular structures and dynamics. In this context, we develop a new optical tomography method working in transmission - Full-field optical transmission tomography (FF-OTT). The method can measure the forward scattering signals and reveal the metabolic time-dependent signals in living cells. FF-OTT is a common path interferometer taking advantage of the Gouy phase shift - a $\pi$ phase shift that the light wave experiences around the focus. By modulating position of the focus one can alter the phase of the scattered light. Demodulation of images with different phases rejects the background and enhances the light from the depth-of-focus, thus producing an optical section. We test FF-OTT by imaging single-cell diatoms and ex vivo biological samples. In fresh samples, we show that the intracellular motions create visible intensity fluctuations in FF-OTT so that the method is able to reveal a metabolic dynamic contrast. FF-OTT was found to be an efficient label free technique that can be readily implemented thanks to a robust common-path speckle-free interferometer design using an incoherent light source.

## 1. Introduction

It is well-known that an object positioned outside of optical microscope focus stays visible and produces a blurred image. This presents a major problem for microscopy: image of interest from a thin depth-of-focus slice can be completely hindered by the superimposed blurred background from the out-of-focus sample volume. With a goal to suppress the background and reveal only the relevant in-focus signals multiple techniques were developed. These are frequently referred to as 'optical sectioning' methods. These methods are numerous and show great diversity in physical and engineering principles that underlie them; therefore, we will restrain ourselves to the most common methods. Optical sectioning methods can be classified in general types: methods physically blocking the out-of-focus light before detection, like confocal microscopy [1], methods filtering the out-of-focus light in post-processing [2], light sheet methods illuminating a single in-focus plane perpendicular to the detection [3], structured illumination methods [4], phase contrast methods highlighting the refractive index changes [5,6], optical coherence tomography methods that obtain cross-sectional [7] and en face sections [8] using the interference of light of low temporal coherence.

One of the earliest and the most widespread optical sectioning method is confocal microscopy (CM) [9], which uses a pinhole mask to physically block the out-of-focus light from reaching the detector. Interestingly, it was noticed that a further improvement in optical sectioning (and axial resolution) can be achieved by looking at the brightness of the scatterer [10]. More precisely, the visible intensity of the scatterer can be linearly correlated to its position by subtracting several images along the point-spread-function (PSF) slope. This approach of optical profiling is called differential confocal microscopy (DSP) and can be also used in a two-camera configuration (one before and one after the focus) [11]. Differential spinning disk (DSD) [4,12] confocal microscopy is another method that relies on two image subtraction, however the images are different by the illumination patterns. Modern confocal systems use a combination of physical (block) and numerical filtering (image subtraction) to improve the background rejection and therefore the optical sectioning.

Although with a different principle, the image subtraction is also used in full-field optical coherence tomography (FF-OCT) [8,13]. FF-OCT uses an interferometer coupled with a light source of low temporal coherence to produce interference between the reference light from the mirror and backscattered light from a thin section of the sample. The phase of the interference is modulated by moving the reference mirror with a piezo-electric motor. Difference between the consecutive phase-shifted images doubles the interference signal and rejects the background thus producing an optical section. FF-OCT works directly in wide-field and does not require point-by-point scanning typical for confocal systems. Another particularity of FF-OCT and OCT techniques in general is that the thickness of the optical section is independent from the depth of field, and can be much smaller than the latter (determined by the spectral bandwidth of the light source). Many applications were found for FF-OCT including biopsy diagnostics [14], visualization of cell organoid dynamics [15,16], in vivo retinal [17] and corneal imaging [18].

Up to now FF-OCT was only used in back-scattering configuration. In the majority of cases confocal microscopy uses a reflection configuration, although an elegant transmission geometry was also proposed [19].

On the opposite, the optical phase contrast methods are primarily of transmission type. Transmission geometry natively enables interference between the light waves scattered by the sample and the waves transmitted without scattering (zero-order diffraction). By using one of the phase contrast techniques such as the classical one from Zernike [5,20], differential interference contrast (DIC) microscopy [21] or defocusing microscopy [22] it is possible to convert phase variations, caused by the light propagating through the thickness of the sample, into the intensity detectable by the camera. The resulting image provides contrasted view into the sub-cellular structures, although the optical sectioning is modest.

In the current paper we introduce a new optical sectioning method that works in transmission. The method takes advantage of the Gouy phase shift [23] – the well-known $\pi$ phase shift that the light wave experiences in the optical focus. The effect of Gouy phase shift modulating the interference signal of freely moving nanoparticles across the focus was shown using coherent [24] and incoherent light sources [25]. Here we study overall still samples, so, instead of relying of the random Brownian motion, we modulate the position of the focus using the piezo-electric motor. Hence, the phase of the scattered light oscillates causing a signal variation. Because the Gouy phase shift is localized at the focus of the objective, only the in-focus light is modulated while the out-of-focus light stays unchanged. Processing of images with different phases rejects the background and enhances the light from the depth-of-focus, thus producing an optical section. In this way the optical sectioning is achieved by a combination of physical and numerical rejection of out-of-focus light (small depth-of-field of focusing optics combined with processing of phase-shifted images). In order to emphasize the transmission nature of the method and the fact that it works in full-field, we call it full-field optical transmission tomography (FF-OTT). We provide a basic theoretical framework for FF-OTT. Then we test its performance by imaging single-cell diatoms and ex vivo samples. Finally, we demonstrate the ability of FF-OTT to image the metabolic dynamic signals in living cells.

## 2. Basic model of full-field optical transmission tomography

*2.1 Principle of FF-OTT*

The Gouy phase shift (GS) is a well-known π phase shift that a converging wave experiences as it passes through its focus. GS is a general property of all waves (light, sound, etc) and is applicable to any converging/diverging waveform [26]. Despite its widespread nature, the possible physical origins of GS were proposed only recently, rooting back to the fundamental uncertainty principle [27] and Huygens principle [28].

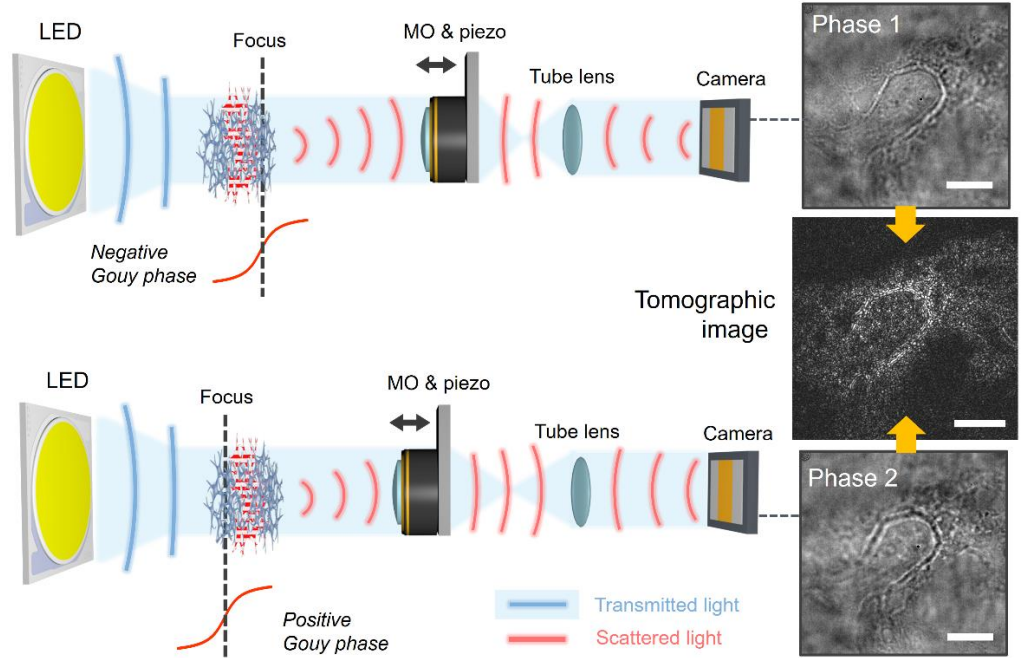

Fig. 1. Principle of FF-OTT: FF-OTT is based on interference of an illuminating wave and a wave scattered by the sample. Shifting position of the microscope objective (MO) modulates the Gouy phase of the scattered wave. Subtraction of the 2 phase shifted images removes the out-of-focus background, thus producing the tomographic image. Sample is a stromal keratocyte cell with visible nuclei from the ex vivo cornea. Scale bars are 10 μm.

Phase changes are not visible to the conventional light detectors that measure intensity. However, one can convert phase variations into intensity variations using interference of the transmitted light wave with the wave scattered by the sample. The transmission interferometer is shown in Fig. 1. Light from the spatially incoherent source such as light-emitting diode (LED) is sent onto the sample. Part of the incoming wave is scattered by the sample and part is directly transmitted though the sample (zero order diffraction) without scattering. The transmitted wave passes through the microscope objective, focusing close to its back focal plane, and gets projected onto the camera by the tube lens. On the opposite, the scattered wave is a diverging wave that is collected by the numerical aperture (NA) of the microscope objective, thus exhibits the GS. The scattered wave gets focused on the camera and interferes with the transmitted light. In order to modulate the interference intensity of the scatterer and thus the location of the scatterer relatively to the focus, the microscope objective is mounted on the piezo-electric motor. Not every scatterer of the sample will exhibit sufficient GS, but only the scatterers within the DOF of the microscope objective (see the simulation below). By processing the images corresponding to different phases, one can reject the out-of-focus background and recover the optical section – the so-called FF-OTT tomographic image. Below,

FF-OTT is analyzed theoretically using a paraxial Gaussian beam approximation and experimentally by imaging nanoparticles.

*2.2 Simulation and experiment with still nanoparticles*

Basically, in our experiments the incoming wave illuminates the sub-wavelength scatterers such as the organelles in a cell volume generating a scattered field that is able to interfere with the incoming wave. We analyze the experiment using the Gaussian beam model, that provides the particularly simple equations.

The electric field of the scattered wave can be written as [29]:

$$E_S(r,z) = E_{S0} \cdot \frac{w_0}{w(z)} \cdot \exp\left(-\frac{r^2}{w^2(z)}\right) \cdot \exp\left(-i \cdot \left(k \cdot z + \underbrace{\frac{k \cdot r^2}{2 \cdot R(z)}}_{\text{Defocus}} - \underbrace{\arctan\left(\frac{z}{z_0}\right)}_{\text{Gouy shift}} - \frac{\pi}{2}\right)\right). \quad (1)$$

Here we for simplicity we assumed an electric field with polarization in r direction and propagation in z direction. In the formula:

r is the lateral distance from the optical axis (r = 0 on the optical axis);

z is the axial distance from optical focus (z = 0 at the focus);

$w_0 = \lambda / \left(\tan\left(\arcsin\left(NA/n\right)\right) \cdot n \cdot \pi\right)$ is the beam waist radius, NA is the numerical aperture;

$z_0 = \pi \cdot w_0^2 \cdot n / \lambda$ is the Rayleigh length, essentially equivalent to half of the depth-of-field.

$E_{S0}$ is the amplitude of the electric field;

$w(z) = w_0 \cdot \sqrt{1 + (z/z_0)^2}$ is the radius, at which the field amplitude falls to 1/e from an on-axis value;

$R(z) = \frac{z^2 + z_0^2}{z}$ is the radius of curvature of the beam's wavefront at z;

$k = 2\pi n/\lambda$ is the wave number, $\lambda$ is the wavelength and n is the refractive index of the medium;

$\pi/2$ is a phase shift due to scattering that is well-known in phase contrast microscopy [30].

On the optical axis at the center of the Gaussian beam (r = 0):

$$E_S(0,z) = E_{S0} \cdot \frac{w_0}{w(z)} \cdot \exp\left(-i \cdot \left(k \cdot z - \arctan\left(\frac{z}{z_0}\right) - \frac{\pi}{2}\right)\right) \quad (2)$$

At the same time, the transmitted wave propagating along the axis follows a simple plane-wave equation:

$$E_T(0,z) = E_{T0} \cdot \exp(-i \cdot k \cdot z) \quad (3)$$

The two fields interfere producing intensity detected by the camera:

$$I_{total} = |E_S|^2 + |E_T|^2 + 2 \cdot \text{Re}\{E_S \cdot E_T^*\} \quad (4)$$

The real interference term:

$$\text{Re}\{E_S \cdot E_T^*\} = E_{S0} \cdot E_{T0} \cdot \frac{z/z_0}{1 + (z/z_0)^2} \quad (5)$$

Here we opened up w(z) and used a trigonometry relation.

In case of a small scattering particle, the interference term is hindered by a strong homogeneous background, and the scattered intensity is neglectable. The goal of FF-OTT is to remove the homogeneous background and isolate the interference term only. Around the focus, where $z/z_0 \ll 1$ the expression of the interference term can be simplified:

$$\text{Re}\{E_S \cdot E_T^*\} = E_{S0} \cdot E_{T0} \cdot \frac{z}{z_0} \tag{6}$$

Hence, close to the focus the interference term evolves linearly with the axial distance to the focus, opening a possibility for tomography. The tomographic image is typically reconstructed by subtracting two images with a slightly different focus in this linear regime. Therefore, the background signal disappears and the tomographic signal turns:

$$I_{tomo} = I_{total1} - I_{total2} = E_{S0} \cdot E_{T0} \cdot \left[\frac{z}{z_0} - \frac{z-p}{z_0}\right] = E_{S0} \cdot E_{T0} \cdot \frac{p}{z_0} \tag{7}$$

with p being the piezo shift. It is interesting to note that the tomographic intensity can be increased by maximizing the piezo shift between the two images, as long as we stay within the linear part around the focus. A general rule of thumb, confirmed by a simulation, is that the linear region roughly corresponds to the half of the depth-of-field.

We simulated Eq. (5) using parameters: NA = 0.5, n = 1.51 (immersion oil), λ = 455 nm. The results are shown in Fig. 2A. One can see that Gouy phase affects the intensity of the interfering waves. Intensity experiences an abrupt linear change within half of the DOF = $z_0$ = 1.5 µm and shows only a limited change outside of focus. The simplest way to get optical sectioning consists in subtracting the two successive images with the different phases leading to enhanced signal from the area around focus and suppression of the background. Optical sectioning effect is illustrated in Fig. 2B. Sectioning can be achieved with any phase modulating step within the linear region. Smaller modulating step ($< z_0$) within this region improves optical sectioning by stronger suppression of the out-of-focus signals. On another hand the larger modulating step ($> z_0$) provides limited suppression of the background signals near the focus, as can be seen by the tails of the curve in Fig. 2B.

As an experimental test for this model, we recorded an axial point-spread-function (PSF) of a small dielectric $TiO_2$ particle (diameter < λ). At the same time, the size was large enough (100 nm) to ensure a sufficient optical signal. The experiment used 100X oil immersion ($n_{oil}$ = 1.51) microscope objective (AmScope, PA100X-INF-IRIS) with NA set to 0.5. The lateral resolution of the system ($2w_0$ = 0.5 µm) was greater than the particle size. The camera was oversampled (1 pixel correspond to 0.07 µm), giving us the possibility to quantitatively explore the particle profile and brightness. Position of the objective was modulated with a piezo motor (NV 40/3 CLE, piezosystem Jena). Illumination from the blue (455nm) LED (M455L4, Thorlabs) was detected by the camera (PhotonFocus, MV-D1024E-160-CL-12) with high full well capacity of 200 000 e-. Results are shown in Figs. 2C, 2D and in **visualization 1**. One can see that intensity, and therefore the brightness of the particle, changes depending on whether it is located before or after the focus. By moving the piezo motor with 100 nm step we get an intensity curve on the particle that resembles the theoretical one in Fig. 2A. At the same time the intensity of the background stays largely unchanged.

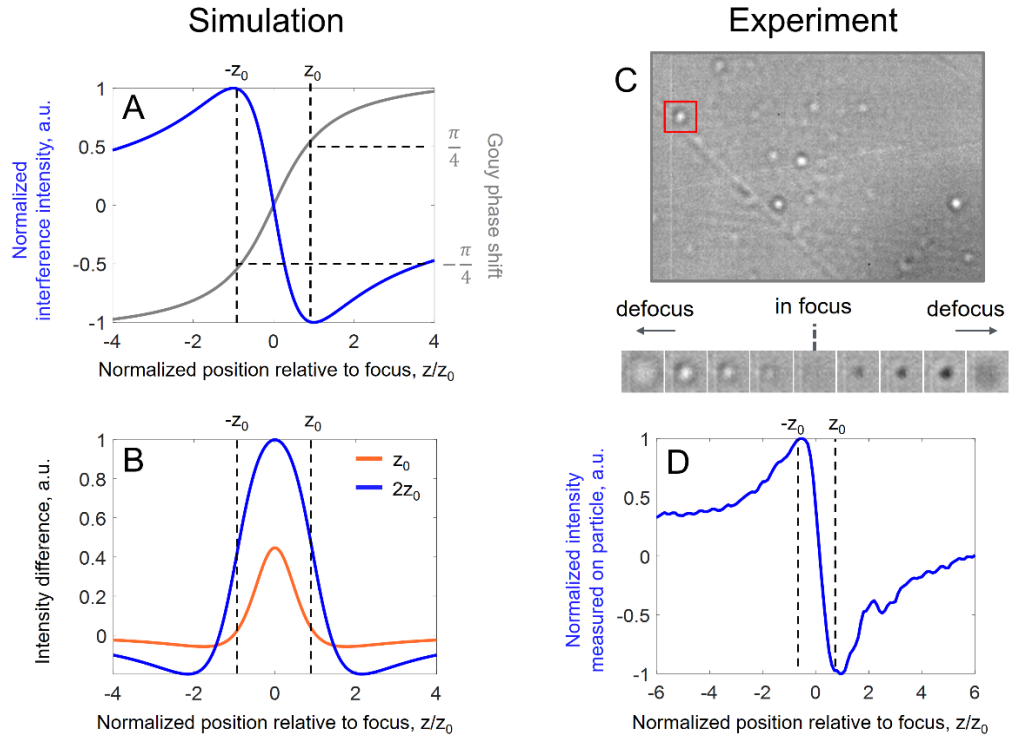

Fig. 2. FF-OTT simulation and experiment with nanoparticles. A: Gouy phase shift and its effect on the intensity of interference light in the focus. DOF = $2z_0$ = 1.5 µm. B: Optical sectioning effect depending on the phase step of the piezo motor. The curves are obtained by subtracting two phase-shifted blue curves in A. C: image on the camera with defocused nanoparticles visible. See the video in **visualization 1**. D: Experimental change of nanoparticle intensity depending on the position relatively to the optical focus (generated from **visualization 1**).

It is interesting to note that the effect of 'transparent objects showing light or dark contours under the microscope in different ways varying with change of focus' [5] was known long before Zernike used it to create a phase contrast microscope in the 1930s [20]. More recently this effect was used in electron microscopy [31] as well in optical microscopy of transparent homogeneous samples [22]. For such objects the wave curvature (defocus) was sufficient to explain the phase shift between the transmitted and scattered waves. Here, like in OCT, we explore objects that, due to their internal heterogenous microstructures, scatter light. Our study suggests that for such micro-scattering objects the Gouy shift might also play a role in phase contrast of these methods. For complex samples the intensity curve will result from the interplay of the Gouy and defocus effects.

## 3. Experimental results

### 3.1 Diatoms

Diatoms are single-cell algae that are of importance for environmental studies. Extended discussion about biology of diatoms can be found in [32] and in the references therein. We used an experimental configuration similar to the mentioned above. An optical slice can be retrieved by subtracting the two phase shifted images. Alternatively the objective can be moved sinusoidally, as in the integrated-bucket FF-OCT [33] in order to perform a lock in detection and increase the sensitivity to small scatterers. Each image was acquired in 3 ms camera exposure time at 135 frames per second, resulting in 15 ms acquisition time for one tomographic frame. Results are shown in Fig. 3. FF-OTT suppresses the fuzzy background that is present in the direct camera images.

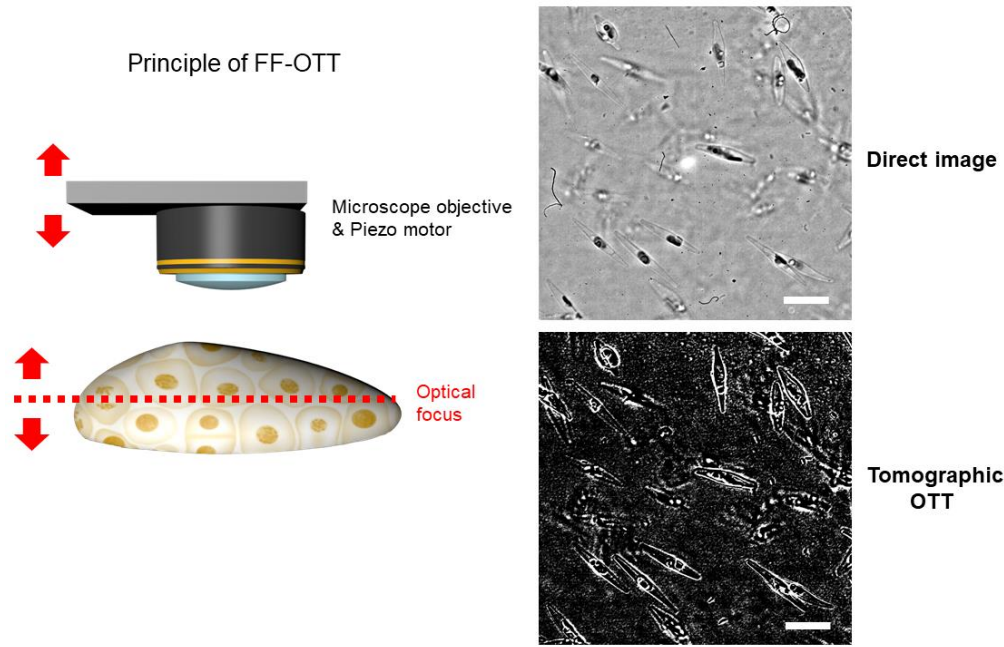

Fig. 3. FF-OTT in the study of diatoms. Direct image refers to a transmission image acquired by the camera without additional processing. Tomographic OTT image was reconstructed from two direct images shifted by a half micrometer. The scale bars are 15 μm.

## 3.2 Ex vivo cornea

Cornea is the most forefront transparent tissue that constitutes the eye. Its exploration is of great interest for medicine as it is a transplantable tissue. Multi-layer composition of the cornea makes its exploration with tomographic techniques particularly interesting and informative.

In order to explore the cell organization, we used an optical system with a smaller magnification but a larger frontal distance. The microscope objective was water immersion ($n_{water}$ = 1.33) 60X NA = 1.1 (NA = 0.5 filled) (LUMFI, Olympus). The theoretical lateral and axial resolutions were 0.6 micron and 1.4 µm, while the FOV was about 180 µm. The samples - ex vivo porcine and macaque corneas were obtained from the partner research institution (Institut de la Vision, Paris) as recuperated waste tissue from an unrelated experiment. They were dissected from the ocular globes within the two hours post-mortem and were imaged within the same day using FF-OTT. In order to limit the spherical aberration due to refractive index mismatch, we first imaged the epithelium and stroma at the front and then flipped the cornea to view the endothelium and stroma at the back.

Corneal images revealed different structures including epithelial cells (40 µm) with nuclei (10 µm), stromal keratocyte cells with nuclei (15 µm) and endothelial cell mosaic (20 µm). The structures showed similar dimensions as visible in FF-OCT and confocal microscopy [18,34,35].

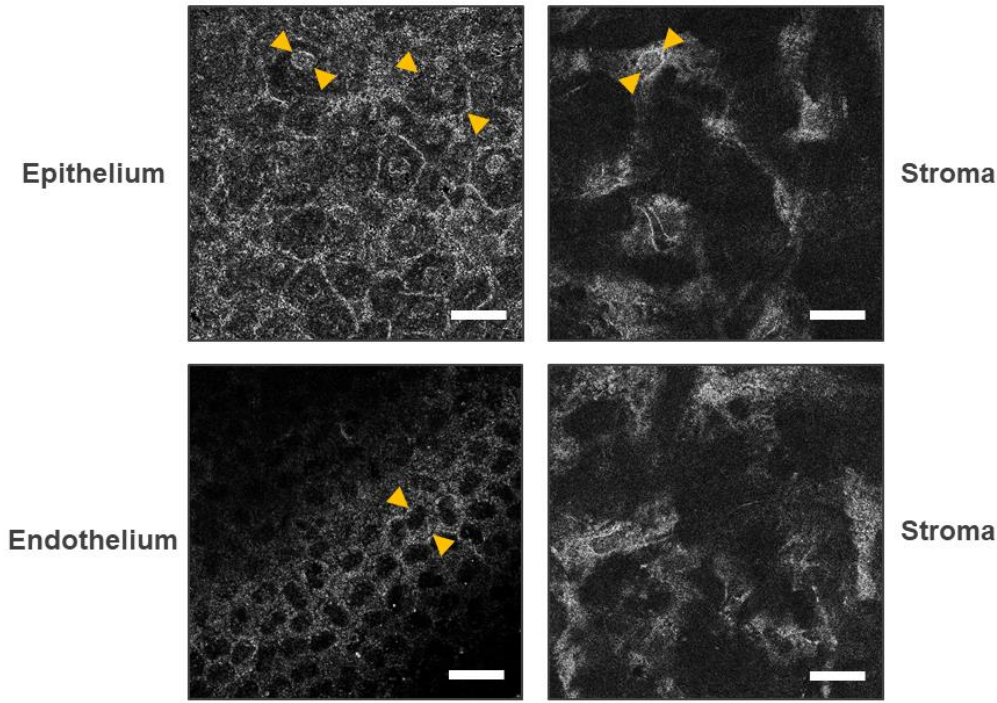

Fig. 4. FF-OTT images of ex vivo human cornea. The stromal and endothelial images were acquired from the macaque cornea, while the epithelial image was taken from the porcine sample (epithelium in the macaque cornea was deteriorated). Arrows highlight the epithelial cell nuclei (10 µm), epithelial cell cytoplasm (40 µm), stromal keratocyte cell nuclei (15 µm) and endothelial cell (20 µm). Sectioning through videos are available as **visualization 2** and **visualization 3**. Scale bars are 30 µn.

### 3.3 Dynamic FF-OTT

One peculiar feature of FF-OTT is that it can be also used to study the metabolic dynamics of the cells. In this context, instead of moving the microscope objective one can rely on the natural movements of the sub-cellular organelles. Thanks to the linearity of the axial interferometric response in FF-OTT, the axial displacements of the live scatterers inside the cell are transferred into the intensity differences. The intensity fluctuation images contain information that can be used to compute the optical section as before, but also contain information to evaluate the active transport of the scatterers inside the cell. Because the intensity strongly depends on the axial response only around the focal plane, the dynamic FF-OTT signal also exhibits optical sectioning. As such this method is a transmission analogue of another label free method - dynamic FF-OCT that is being used in back-scattering [15,16].

An extended analysis of dynamic FF-OTT application to diatoms can be found in a biology-oriented paper [32], while in this work we demonstrate the promise of the method for imaging 2D cell cultures and 3D retinal layers.

We first illustrate the dynamic FF-OTT method by studying the behavior of HeLa cells (93021013-1VL, Merck) [36] grown on a cover slip in a Petri dish. The cells were positioned in the best focus of FF-OTT device. Here we used a high resolution 100X oil immersion FF-OTT device mentioned above. FF-OTT acquired a time lapse stack of 256 direct camera images in 3 seconds. Brightness of each pixel of the stack fluctuates with a certain average frequency and frequency bandwidth, reflecting the movements of the organelles located in that pixel. By assigning each of these variables to a respective channel in the HSV color map (H - avg. frequency, S – frequency bandwidth, V – brightness), one can create a colored dynamic FF-OTT image (Fig. 5). In these images, the red color corresponds to the fast frequencies, thus the fast-moving organelles.

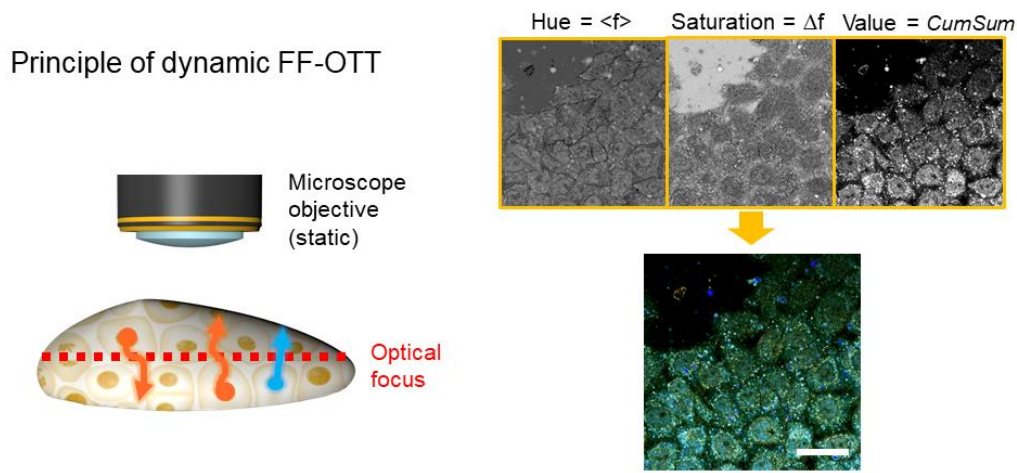

Fig. 5. Dynamic FF-OTT principle. Instead of moving the optical focus, one can rely on natural movements of sub-cellular organelles around the optical focus for phase shifting. The stack of 256 direct images contains sufficient information to reconstruct a tomographic image with the cell metabolism encoded in the color (blue = slow, red = fast). This is done by looking at the time evolution of each pixel of the stack and by measuring the average frequency of modulation <f.>, frequency range Δf and cumulative sum (CumSum). Then HSV image is created by assigning these parameters to each channel. Scale bar is 20 μm.

In order to confirm that the color signal originates from the cell metabolism we applied on the sample the D-oxy-glucose that inhibits glycolysis and stops part of the energy supply. The evolution of dynamic FF-OTT images is shown in Fig. 6. As expected the cytoplasm signal decreases in time [37].We also noticed the 2 to 5 times increase in dynamic signal from the nuclei. In order to better characterize this dynamics we compared the standard deviation and the cumulative sum, following the method developed by Scholler [37]. We found that the normalized cumulative sum was about the same as the standard deviation (ratio 1.3) before addition of the D-oxy-glucose and it increased to ratio 2.5 after the addition. This result suggests that the random movement of scatterers within the nuclei is likely to be associated to a drift that makes it anomalous, rather than being a pure Brownian motion. Indeed, anomalous movements have been already described in the nucleus and D-oxy-glucose has been shown to inhibit protein transport from cytoplasm to nucleus [38,39].

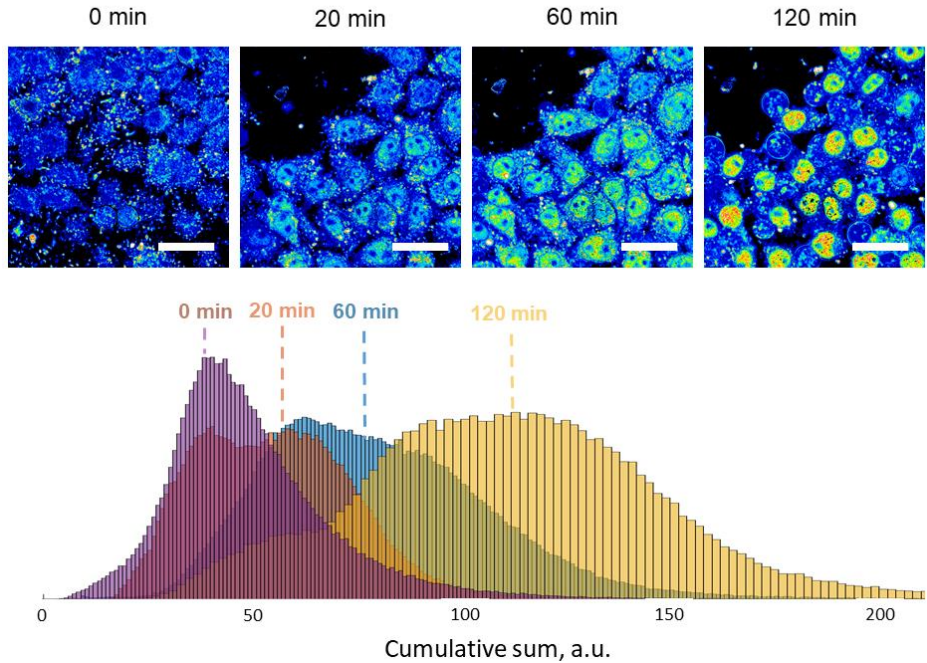

Fig. 6. Dynamic FF-OTT application to imaging HeLa cell metabolism. Introduction of D-oxy-glucose partly blocks the cell energy supply resulting in the expected reduced dynamic signal from the cytoplasm and increased signal from the nuclei, as seen by FF-OTT. Histogram is calculated by measuring the cumulative sum signal at each pixel within the nuclei. The method did not require any fluorescent biomarkers. Scale bars are 20 μm.

Dynamic FF-OTT can be performed not only in cell cultures, but also in thick samples like ex vivo pig retina. The results are shown in **visualization 4**. The ex vivo pig retina was obtained from the partner research institution (Institut de la Vision, Paris) as recuperated waste tissue from a regional slaughterhouse. The retina was dissected from the ocular globe within two hours post-mortem and was imaged within the same day using FF-OTT. The tissue was put on a simple coverslip and was immerged in PBS. The drift in the liquid was causing occasional motion artifacts in the images. The light traveled through the full thickness of the retina, while the ganglion cell layer, close to the surface of the retina was imaged. 256 images were acquired at each plane, separated by 2 μm, and one dynamic FF-OTT image was calculated at each plane. The direct intensity image, the static FF-OTT and the dynamic FF-OTT images were combined together **in visualization 4**. The first layer of flat and large cells, possibly glial cells is visible in dynamic FF-OTT, followed by a few planes, where large and small circular ganglion cells as well as their nuclei can be observed together with large axons in static FF-OTT. The retinal ganglion cells exhibit smaller amplitude but faster dynamics as compared to the first class of (blue) cells. Finally, the inner plexiform layer with many point-like synapses was visible, together with the first few bipolar cells from the inner nuclear layer. Altogether, many different cells and features of the retina can be investigated by combining static and dynamic FF-OTT.

## 4. Discussion, conclusion and perspectives

FF-OTT introduces a new optical sectioning method that exploits the effect of Gouy phase shift. The optical sectioning of about depth-of-field is achieved by a combination of the physical (high numerical aperture) and numerical (phase shifting) rejection of out-of-focus light. One of the main advantages of the method is the particularly simple setup. The common-path interferometer design with a single microscope objective does not require high precision mechanics and is immune to mechanical/thermal misaligning. FF-OTT can be used to get a distribution of microorganisms (e.g. algae) and to section through ex vivo samples. The method is particular interesting for imaging ex vivo cornea samples, as it shows similar features as FF-OCT and confocal microscopy, while using a much simpler device. As such, FF-OTT is promising for diagnostics of corneal grafts.

Among the limitations of the method, we should mention the finite damping of the out-of-focus light (see the wings of the curve in Fig. 2B), when compared to the strong interferometric damping in OCT. Indeed, the OCT sectioning (axial PSF) is usually linked to the Fourier transform of the close to Gaussian source spectrum and quickly converges to zero. For FF-OTT, as could be seen on figure 2B, we need to use an axial modulation smaller than half the depth of field to get an acceptable damping of the axial PSF wings.

Comparing to FF-OCT, the axial resolution in the new method is determined by the numerical aperture of the optics used and not the spectral bandwidth of the light source, which limits FF-OTT to the use of relatively high NA objectives. Another distinctive feature of FF-OTT is that it uses a common-path interferometer, which eliminates the problem of mismatch between the optical focus and the coherence plane present in FF-OCT [40]. However, the lack of independent reference arm means that both the sample and reference signals are getting damped, when propagating through thick strongly absorbing or scattering media. Part of this disadvantage is compensated by the fact that biological tissues are mostly "forward scatterers" due to the size of cells and nuclei (scattering by the structures larger than hundred nanometers is highly anisotropic in the forward direction). Lastly, in contrast to FF-OCT the new method does not produce the optical interference fringe artifacts typically observed in cells cultures on glass or plastic slides, or in regular cell mosaics, such as corneal endothelium [41].

The most peculiar feature of FF-OTT is its ability to visualize the metabolic cell dynamics without fluorescent labels. In this sense, FF-OTT can produce images similar to the dynamic backscattering-based FF-OCT [15,16], while using a much simpler and robust optical design.

Although our simple theoretical model shows agreement with experiment, the understanding of FF-OTT would greatly benefit from a more rigorous theoretical analysis in the future. This model should take into account the high-NA optics (not completely valid for Gaussian beams) as well as the different numerical apertures for the transmitted and scattered waves. More precisely, the NA of the transmitted beam is determined by the LED illumination angle, while the NA of the scattered wave depends on the size of the scatterers. Finally, the scattering from the real samples is complex - even the scattering electric field from the 100 nm particles is located on a larger region than the particle itself. The angular distribution as well as the geometrical parameters of the scattered electric field should be taken into account.

**Funding.** This work was supported by the PSL pre-maturation grant under the French Government program "Investissements d'Avenir" (ANR-10-IDEX-0001-02 PSL), the ANR PRIMAVERA, the proof of concept (POC) grant funded by the European Research Council (ERC) (957546), the recurrent budget of the Institut Langevin Langevin (CNRS and ESPCI). Samer Alhaddad is beneficiated of a PhD grant funding from the European Union's Horizon 2020 research and innovation program under the Marie Skłodowska-Curie grant agreement No 754387.

**Acknowledgments.** We warmly thank Morgane Corre from IBENS for the HeLa cells cultures on glass slides that we used in this study.

**Disclosures.** OT: (P), MB: (P), ACB (P), SA: none, VM: none.

**Data availability.** Full resolution images generated in this paper are not publicly available at this time but may be obtained from the authors upon reasonable request.

**Supplemental document.** See Visualization 1, Visualization 2, Visualization 3, and Visualization 4 for supporting content.